\documentstyle[preprint,aps,epsf,floats,rotate]{revtex}                     
\tighten
\begin{document}

\preprint{
\vbox{\hbox{CALT-68-2154}
      \hbox{UTPT-97-24}
      \hbox{hep-ph/9801328} } }

\title{Photoproduction of $h_c$}
\author{Sean~Fleming\footnote{{\tt fleming@furbaide.physics.utoronto.ca}}}
\address{ Physics Department, The University of Toronto
 60 St.~George St., Toronto, Ontario  M5S~1A7, Canada }
\author{Thomas~Mehen\footnote{{\tt mehen@theory.caltech.edu}}}
\address{California Institute of Technology, Pasadena, CA 91125, U.S.A.}

\date{January 1998}

\maketitle
\begin{abstract}

Using the NRQCD factorization formalism, we calculate the total cross 
section for the photoproduction of $h_c$ mesons. We include color-octet 
and color-singlet mechanisms as well as next-to-leading order perturbative 
QCD corrections. The theoretical prediction depends on two nonperturbative 
matrix elements that are not well determined from existing data on charmonium 
production. For reasonable values of these matrix elements, the cross section 
is large enough that the $h_c$ may be observable at the E831 experiment and 
at the HERA experiments.

\end{abstract}

\pagebreak 

Of all the charm-anticharm quark boundstates lying below the threshold
of open charm production, the most elusive to experimental
investigation is the $h_c$ meson. To date this particle has been 
observed by only a few experiments, which measure an average mass
of $M_{h_c} = 3.52614 \pm 0.00024 \; \mbox{GeV}$~\cite{e760}. 
Studies of the 
$h_c$ are difficult because it has quantum numbers $\rm {J^{PC}} = 1^{+-}$, 
and thus cannot be produced resonantly in $e^+ e^-$ annihilation, or 
appear in the decay of a ${\rm J^{PC}} = 1^{--}$ charmonium state via an 
electric dipole transition.

Since charm quarks are heavy compared to $\Lambda_{\rm{QCD}}$, it is
natural to view charmonium as a nonrelativistic system, where the
$h_c$ is a spin-singlet P-wave state of a $c$ and $\bar{c}$. This
approach is taken in the color-singlet model~\cite{schuler} (CSM) of
quarkonium production and decay. In the CSM, it is assumed that the
$c\bar{c}$ must be produced in a color-singlet state with the same
angular-momentum quantum numbers as the charmonium meson which is
eventually observed.  However, the CSM has serious deficiencies. It is
well known that perturbative QCD calculations of production and decay
of P-wave quarkonia within the CSM are plagued by infrared
divergences~\cite{barb}. More recently, measurements made at the
Fermilab Collider Detector Facility (CDF) show that the CSM also fails
to accurately predict the production cross sections of S-wave
quarkonia ($J/\psi$ and $\psi'$) \cite{sansoni}.

The naive CSM has been supplanted by the NRQCD factorization formalism
of Bodwin, Braaten, and Lepage~\cite{bbl}. This formalism allows the
infrared safe calculation of inclusive charmonium production and decay
rates. It also predicts new mechanisms in which a $c\bar{c}$ pair is
produced at short distances in a color-octet state, and hadronizes
into a final state charmonium nonperturbatively. These color-octet
mechanisms can naturally account for the CDF data on $J/\psi$ and
$\psi'$ production.

In this paper, we examine photoproduction of $h_c$ within the NRQCD
factorization formalism.  Color-octet mechanisms play an essential
role. By itself the CSM yields infrared divergent expressions for the
$h_c$ cross section; however, once color-octet contributions are
included we obtain sensible predictions. The resulting expression then
depends on two undetermined nonperturbative matrix elements. Using
heavy-quark spin symmetry these can, in principle, be estimated from
similar matrix elements extracted from data on the production and
decay of $\chi_{cJ}$ mesons. Currently, theoretical and experimental
uncertainties in the determination of the $\chi_{cJ}$ matrix elements
preclude us from making definitive predictions for the $h_c$
photoproduction cross section. This will be discussed in greater
detail below. However, for some choices of the matrix elements which
are consistent with current data, we find a large enough cross section
that the $h_c$ may be observed in current photoproduction experiments
at the DESY HERA collider and at the Fermilab fixed target experiment
E831.

In the NRQCD factorization formalism, inclusive quarkonium production
cross sections have the form of a sum of products of short-distance
coefficients and NRQCD matrix elements.  The short-distance
coefficients are associated with the production of a heavy
quark-antiquark pair in specific color and angular-momentum states.
They can be calculated using ordinary perturbative techniques, and are
thus an expansion in the strong coupling constant $\alpha_s$.  The
NRQCD matrix elements parameterize the hadronization of the
quark-antiquark pair, and each scales as a power of the average
relative velocity $v$ of the heavy quark and antiquark as determined
by the NRQCD velocity-scaling rules~\cite{lmnmh}.

The NRQCD factorization formula for $h_c$ photoproduction, at leading
order in $v$, is
\begin{eqnarray}
  \label{nrqcdcs}
\sigma(\gamma + N \to h_c + X) & = & \int dx \; \sum_i f_{i/N}(x) \; 
\left[ 
\hat{\sigma}(\gamma + i \to c\bar{c}({\bf 8},{}^1S_0) + X) 
\langle {\cal  O}^{h_c}_8(^1S_0)\rangle^{(\mu)} +
\right. 
\nonumber \\
&& \;\;\;\;\;\;\;\;
\left.
\hat{\sigma}(\gamma + i \to c\bar{c}({\bf 1},{}^1P_1) + X; \mu) 
\langle {\cal  O}^{h_c}_1(^1P_1)\rangle \right] \; , 
\end{eqnarray}
where $f_{i/N}(x)$ is the probability of finding a parton $i$ in the
nucleon with a fraction $x$ of the nucleon
momentum. $\hat{\sigma}(\gamma + i \to c\bar{c}({\bf 8},{}^1S_0) + X)$
is the short-distance coefficient for producing a $c \bar{c}$ pair in a
color-octet ${}^1S_0$ configuration, and 
$\hat{\sigma}(\gamma + i \to c\bar{c}({\bf 1},{}^1P_1) + X;\mu)$ is the
short-distance coefficient for producing a  $c \bar{c}$ pair in a
color-singlet ${}^1P_1$ configuration.
The matrix element $\langle {\cal  O}^{h_c}_{1(8)}(^{2S+1}L_J)\rangle$
describes the hadronization of a color-singlet (color-octet) 
${}^{2S+1}L_J$  $c\bar{c}$ pair into an $h_c$. According to the 
$v$-scaling rules, both matrix elements in Eq.~(\ref{nrqcdcs}) scale
as $v^5$. The scale $\mu$ appearing in Eq.~(\ref{nrqcdcs}) arises from
the factorization of the cross section into long-distance and
short-distance contributions. As we discuss below, the $\mu$ dependence
of the color-octet matrix element is cancelled by the $\mu$ dependence
of the color-singlet short-distance coefficient so that the expression
for the cross section is $\mu$ independent. Note that in addition to
the NRQCD scale $\mu$ shown explicity in Eq.~(\ref{nrqcdcs}) there is
a renormalization scale and a factorization scale (associated with the 
parton distribution function $f_{i/N}(x)$) which have been suppressed. 
These three scales do not have to be the same, however, in our numerical 
calculations we choose them to be equal.

The short-distance coefficients in Eq.~(\ref{nrqcdcs}) can
be calculated using the techniques of Ref.~\cite{bc12}. At leading
order in $\alpha_s$, we obtain the following expression for the
color-octet short-distance coefficient:
\begin{equation}
  \label{octet}
  {d \hat{\sigma} \over dz}
  (\gamma + g \to c\bar{c}({\bf 8}, {}^1S_0))= 
  {\pi^3 \alpha \alpha_s e^2_c \over 4 m^5_c} \; 
  \delta(1-z) \; ,
\end{equation}
where $z=\rho/x$, $\rho = 4m^2_c/S_{\gamma N}$, and $S_{\gamma N}$ is the
photon-nucleon center-of-mass energy squared. We calculate the color-singlet
cross section, regulating the infrared divergences using dimensional
regularization. The result in the $\overline{\mbox{MS}}$ scheme is:
\begin{eqnarray}
  \label{singlet}
\lefteqn{
  { d\hat{\sigma} \over dz}
  (\gamma + g \to c\bar{c}({\bf 1}, {}^1P_1) + g; \mu) =
  {8 \pi^2 \alpha \alpha^2_s e^2_c \over 27 m^7_c} \;
  \left[ f(z)+ 
  {z^4 (1+z^2)\over (1+z)^2} {1 \over (1-z)_{\rho}}
\right.}
\nonumber \\
&& \;\;\;\;\;\; \left.
+ \; 2 \; \delta(1-z)\left( \ln(1-\rho) - {5 \over 6} \right) \right] - 
{\pi^3 \alpha \alpha_s e^2_c \over 4 m^5_c}\; \delta(1-z)
\left[
{16 \alpha_s \over 27 \pi m^2_c} \ln\left({\mu \over 2m_c}\right)
\right]  \; ,
\end{eqnarray}
where
\begin{eqnarray}
  \label{fz}
  f(z) &=& -{z^5 \ln(z) \over (1+z)^3} +
 {z^2 (5+3z+14z^2+2z^3+9z^4-z^5)\ln (z)  \over
   (1-z)^3 (1+z)^5 }     
\nonumber \\
&& \;\;\;\;\;\;
+ {z^2 (1+z+10z^2+4z^3+15z^4-z^5+2z^6) \over
      (1-z)^2(1+z)^5} \; ,
\end{eqnarray}
and the functional distribution is defined by
\begin{equation}
  \label{roplus}
  \int^1_{\rho} dz \; f(z) \left( {1 \over 1-z} \right)_{\rho} = 
\int^1_{\rho} dz \; { f(z) -f(1) \over 1-z} \; .
\end{equation}

The NRQCD expression for the cross section is obtained by substituting
Eqs.~(\ref{octet})~and~(\ref{singlet}) into Eq.~(\ref{nrqcdcs}).  
Note that a $1/\epsilon$ divergence in the color-singlet coefficient
has been absorbed into the definition of the leading color-octet
matrix element. The renormalized matrix element 
$\langle{\cal O}^{h_c}_8(^1S_0)\rangle^{(\mu)}$ can be shown to obey
the renormalization group equation~\cite{bbl},
\begin{equation}
  \label{rge}
\mu  { d \over d \mu}  \langle {\cal O}^{h_c}_8(^1S_0)\rangle^{(\mu)}
= {16 \alpha_s \over 27 \pi m^2_c} \langle {\cal O}^{h_c}_1(^1P_1)\rangle
\; .
\end{equation}
Thus the logarithmic dependence on $\mu$ of the short-distance
coefficient 
$\sigma(\gamma + g \to c\bar{c}(\mbox{\bf{1}},{}^1P_1 + X; \mu)$
is canceled to this order by the $\mu$ dependence of the renormalized
matrix element $\langle{\cal O}^{h_c}_8(^1S_0)\rangle^{(\mu)}$.

Note that the leading color-octet coefficient is $O(\alpha_s)$ while
the leading color-singlet coefficient is $O(\alpha^2_s)$. Therefore,
next-to-leading order QCD corrections to the color-octet coefficient
are of the same order as the leading color-singlet coefficient, and
must be included if we are to have a complete $O(\alpha^2_s$)
calculation of $h_c$ photoproduction. The $O(\alpha^2_s)$ corrections
to the color-octet contribution are computed in Ref.~\cite{mmp}. These
corrections are included in our calculation of the cross section.

In order to make a prediction for the $h_c$ photoproduction cross
section, we must determine the values of the nonperturbative matrix
elements 
$\langle{\cal O}^{h_c}_8(^1S_0)\rangle^{(\mu)}$
and
$\langle {\cal O}^{h_c}_1(^1P_1)\rangle$. To eliminate large
logarithms in the expression for the cross section we choose
$\mu = 2m_c$. At this time there does not exist a direct
measurement of these matrix elements. However, they are related to
similar matrix elements for $\chi_{cJ}$ production and decay by the
(approximate) heavy quark spin symmetry of NRQCD: 
\begin{eqnarray}
  \label{merel}
\langle {\cal O}^{\chi_{c1}}_1(^3P_1)\rangle &=& 
\langle {\cal O}^{h_c}_1(^1P_1)\rangle \; (1 + O(v^2))
\nonumber \\
\langle{\cal O}^{\chi_{c1}}_8(^3S_1)\rangle &=&  
\langle{\cal O}^{h_c}_8(^1S_0)\rangle \; (1 + O(v^2)) \; .
\end{eqnarray}
The size of the $O(v^2)$ corrections in Eq~(\ref{merel}) can be
estimated by studying radiative $\chi_c$ and $\psi^{\prime}$ decays,
where predictions based on heavy quark spin symmetry agree with
experiment to 20\%  
accuracy~\cite{wc}.

$\langle {\cal O}^{\chi_{c1}}_1(^3P_1)\rangle$ can be extracted from
$\chi_{cJ}$ decays, or from the decay  $B \to \chi_{cJ} + X$. The
authors of Ref.~\cite{mp} calculate inclusive hadronic $\chi_{cJ}$ decay
including next to leading order $\alpha_s$ corrections. The result of
their fit is: 
\begin{equation}
  \label{mp1}
 { \langle {\cal O}^{\chi_{c1}}_1(^3P_1)\rangle \over m^2_c }
= 0.115 \pm 0.016 \; \mbox{GeV}^3 \; . 
\end{equation}
The error includes only experimental uncertainties. It is particularly
important to note that Ref.~\cite{mp} does not include $O(v^2)$
relativistic corrections which are numerically of the same size as the
$O(\alpha_s)$ perturbative corrections that have been
included. Relativistic corrections to the $J/\psi$ decay rate are
large~\cite{gk}. In the case of $\chi_{cJ}$ decay, relativistic
corrections need to be analyzed before 
$\langle {\cal O}^{\chi_{c1}}_1(^3P_1)\rangle$ can be extracted with
confidence. 

The decay $B \to \chi_{cJ} + X$ is calculated in
Ref.~\cite{bbly}. Measurements of $B$ decay~\cite{cleo}
allow an extraction of $\langle {\cal O}^{\chi_{c1}}_1(^3P_1)\rangle$:
\begin{equation}
  \label{bbly1}
 { \langle {\cal O}^{\chi_{c1}}_1(^3P_1)\rangle \over m^2_c }
= 0.42 \pm 0.16 \; \mbox{GeV}^3 \; . 
\end{equation}
Again, the error quoted above is only due to experimental
uncertainties. The authors of Ref.~\cite{bbly} state that their
calculation suffers from large theoretical uncertainties due to next
to leading order QCD corrections to the Wilson coefficients and the
subprocess $b \to c \bar{c} s$. Note that the two extractions of 
$\langle {\cal O}^{\chi_{c1}}_1(^3P_1)\rangle$ agree only at the 
$2 \sigma$ level.

$\langle{\cal O}^{\chi_{c1}}_8(^3S_1)\rangle$ can be extracted from
the decay $B \to \chi_{cJ} + X$, and from CDF data on $\chi_{cJ}$
production. The result of the fit to $B$ decay, after running from the
scale $m_b$ to $2 m_c$ and converting from a cutoff regularization
scheme to dimensional regularization, is
\begin{equation}
  \label{octbdecay}
  \langle{\cal O}^{\chi_{c1}}_8(^3S_1)\rangle^{(2 m_c)}
= (2.9 \pm 2.0) \times 10^{-2} \; \mbox{GeV}^3 \; .
\end{equation}
As before only experimental errors are included. The result of two
different fits to Tevatron data are 
\begin{eqnarray}
  \label{octteva}
  \langle{\cal O}^{\chi_{c1}}_8(^3S_1)\rangle
&=& 0.98 \pm 0.13 \times 10^{-2} \; \mbox{GeV}^3
\\
  \label{octtevb}
  \langle{\cal O}^{\chi_{c1}}_8(^3S_1)\rangle^{(2 m_c)}
&=& 2.6 \times 10^{-2} \; \mbox{GeV}^3 \; ,
\end{eqnarray}
where the values are taken from Refs.~\cite{cl} and~\cite{bfy}
respectively. The error of Ref.~\cite{cl} includes only experimental
uncertainty; Ref.~\cite{bfy} does not quote errors.  The central
values of both fits lie within the $1 \sigma$ error of the extraction
from $B$-decays. 

Note that in the $\chi_{cJ}$ production calculations of Ref.~\cite{cl}, 
there is no color-singlet contribution that gives an infrared
divergence which needs to be absorbed into the definition of the color-octet 
matrix element, as is done in our calculation. Therefore, it is not possible
to relate the the ``bare'' matrix element appearing in
Eq.~(\ref{octteva}) with the renormalized matrix element needed for
our calculation. However, the fragmentation calculation of $\chi_{cJ}$
production carried out in Ref.~\cite{bfy} makes use of the $g \to \chi_{cJ}$
fragmentation function. This fragmentation function includes both
color-octet and color-singlet contributions, and has an infrared
divergence in that is absorbed into the definition of the color-octet
matrix element. This allows us to import the extracted value given in
Eq.~(\ref{octtevb}) into our calculation.

Not all values for $\langle {\cal O}_8^{h_c}(^1S_0) \rangle$ and 
$\langle {\cal O}_1^{h_c}(^1P_1) \rangle$ result in a physically sensible 
prediction for the cross section. Once the infrared divergence from the 
color-singlet contribution to the cross section is factorized, the remaining 
finite contribution is actually negative (for positive 
$\langle {\cal  O}^{h_c}_1(^1P_1)\rangle/m^2_c$). If the 
ratio of the color-octet to color-singlet matrix elements is too small,
it is possible to obtain physically meaningless results.  This is the case 
if, for example, the central values of the color-singlet matrix
element extracted  
from $B$-decays (Eqs.~(\ref{bbly1},\ref{octbdecay})) is used. Therefore it is
impossible to put a lower bound on the cross section given our current
state of ignorance concerning this matrix element. It is also
important to point out that the NRQCD velocity scaling rules imply
that $\langle{\cal O}^{h_c}_8(^1S_0)\rangle^{(2m_c)}$ and
$\langle {\cal O}^{h_c}_1(^1P_1)\rangle/m^2_c$ should be roughly the
same size. Thus, theory would prefer a smaller value for the matrix element
$\langle {\cal  O}^{h_c}_1(^1P_1)\rangle$, as
suggested by the fit to $\chi_{cJ}$ decay.
Clearly more accurate 
extractions from the Tevatron, from $B$-decays, and from $\chi_{cJ}$ decays are needed to clarify 
the situation. This may be possible once next-to-leading perturbative QCD 
corrections and leading relativistic corrections to these processes are 
calculated.

\begin{figure}[htbp]
  \begin{center}
\rotate[l]{
\epsfxsize=10cm
\hfil\epsfbox{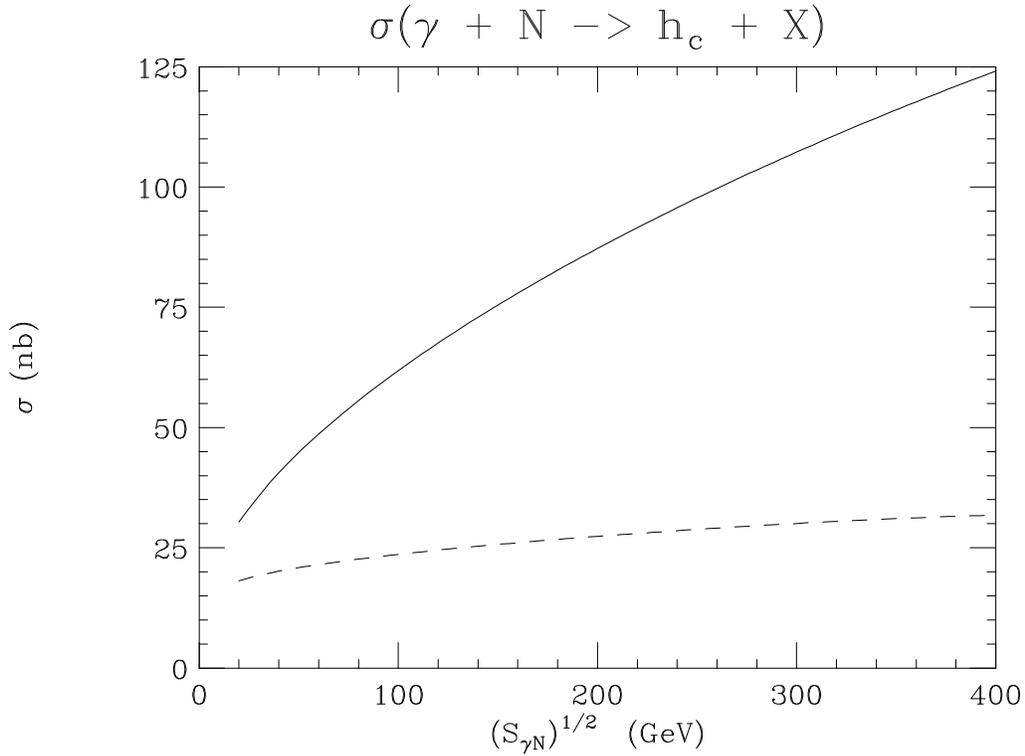}\hfill }    
    \caption{The total $h_c$ photoproduction cross section
$\sigma(\gamma + N \to h_c +X)$ as a function of the center of 
mass energy $\rm \protect \sqrt{S_{\gamma N}}$. The solid curve is for
the choice $\langle {\cal O}^{h_c}_1(^1P_1)\rangle/m^2_c = 0.115 \;
{\rm GeV^3}$ and $\langle{\cal O}^{h_c}_8(^1S_0)\rangle^{(2m_c)} =
2.6 \times 10^{-2} {\rm GeV^3}$. The dashed line is obtained by
keeping the same value for the color-octet matrix element and by 
changing the value of the color-singlet matrix element to
$\langle {\cal  O}^{h_c}_1(^1P_1)\rangle/m^2_c=0.2\; {\rm GeV^3}$. }  
    \label{csvss}
  \end{center}
\end{figure}

For the NRQCD matrix elements, we will use 
$\langle {\cal  O}^{h_c}_1(^1P_1)\rangle/m^2_c=0.115\; {\rm GeV^3}$ as
determined by the 
analysis of $\chi_{cJ}$ decays (Eq.~(\ref{mp1})), and we choose  
$\langle{\cal O}^{h_c}_8(^1S_0)\rangle^{(2m_c)} = 2.6 \times 10^{-2} 
{\rm GeV^3}$. We use the CTEQ~3M parton distribution function with the 
factorization scale choosen to be equal to the NRQCD scale $\mu = 2 m_c$.
The resulting cross section is plotted as the
solid line in Fig.~\ref{csvss}. However, our results are extremely
sensitive to the uncertainty in the determination of the matrix
elements. To show this, we also plot, as the dashed line, the cross
section with the choice  
$\langle {\cal  O}^{h_c}_1(^1P_1)\rangle/m^2_c=0.2\; {\rm GeV^3}$. 
The cross section then drops by roughly a factor of four. 

We conclude with a brief discussion of the possibility of observing
the $h_c$. The E831 photoproduction experiment corresponds to roughly 
$\sqrt{S_{\gamma N}} = 20 \; \mbox{GeV}$, while the HERA experiments
take photoproduction data at approximately 
$\sqrt{S_{\gamma N}} = 100 \; \mbox{GeV}$. At these energies the cross
section for $h_c$ production, assuming our first choice of NRQCD
matrix elements, is 30~nb and 62~nb respectively. These cross sections
are comparable to $J/\psi$ production. The $h_c$ can be detected either
via the rare decay $h_c \to J/\psi + \pi$ with branching ratio on the
order of 1\%~\cite{rare}, or the radiative decay 
$h_c \to \eta_c + \gamma$ (BR = 50\%)~\cite{bbl2}. If it is possible to
reconstruct these decay modes, then the possibility of observing the
$h_c$ is real, and should be experimentally investigated. 

We would like to thank Christian Bauer, Adam Falk,  Martin Gremm, Mike
Luke, and Mark Wise
for helpful discussions. T.M. would also like to thank the Fermilab
summer visitors program for their hospitality while some of this work
was being done. The work of S.F. is supported by NSERC.
The work of T.M. is supported in part by the Department of Energy
under grant number DE-FG03-ER 40701 and by a John A. McCone Fellowship.

\end{document}